\documentclass[twoside,12pt]{article}

\usepackage{amsmath}
\usepackage{amssymb}
\usepackage[dvips]{epsfig}

\newcommand{\be}{\begin{equation}}
\newcommand{\ee}{\end{equation}}
\newcommand{\bea}{\begin{eqnarray}}
\newcommand{\eea}{\end{eqnarray}}

\begin{document}

\title{ \vspace{1cm} Hard Exclusive Photoproduction of
$\Phi$ and J/$\Psi$ Mesons}
\author{A.T.\ Goritschnig$^{1}$, B.\ Meli\'c$^2$,
K.\ Passek-Kumeri\v cki$^2$, W.\ Schweiger$^1$
\\
$^1$\small{Karl-Franzens-University Graz, Institute of Physics,
Division of Theoretical Physics, Austria}\\
$^2$\small{Rudjer Bo\v skovi\'c Institute Zagreb,
Theoretical Physics Division, Croatia}\\
}
\date{}
%\date{20.11.2007(v2)}
\maketitle
\begin{abstract} \noindent
We calculate the leading-order perturbative contribution to
$\gamma\, p \rightarrow M_{V}\, p$, with $M_V$ being a $\Phi$ or
$J/\Psi$ meson, in the kinematic region of large energy and
scattering angle.
\end{abstract}
%\eject
%\tableofcontents
%\section{Introduction}
The theoretical basis of our investigation is the ERBL
factorization scheme for hard exclusive hadronic reactions (see
Refs.~\cite{1,2}). This scheme results from an asymptotic
analysis, which is undoubtedly valid for infinitely large momentum
transfer, but it strongly depends on the particular reaction
whether it can be applied to kinematic situations accessible in
experiments. Only if competing (non-perturbative) mechanisms play
a minor role for one or the other reason, the ERBL contribution
has a fair chance to provide a substantial part of the cross
section even at moderately large momentum transfer. This could be
the case for the $\Phi$ and $J/\Psi$ photoproduction channels, 
since vector-meson-dominance as well as handbag-type mechanisms
are suppressed if a heavy quark-antiquark pair has to be produced.
This is our main motivation to concentrate on the particular
photoproduction channels.

According to the ERBL factorization scheme a hadronic scattering
amplitude $M$ at large momentum transfer can be written as a
convolution integral of a hard scattering amplitude $T_{\mathrm
H}$, describing the scattering of the hadronic constituents, and
hadronic distribution amplitudes (DAs) $\phi_{\mathrm H}$,
parameterizing their bound state dynamics:
\be M_{\gamma p \rightarrow M_V p}\, (s,t) = \int_0^1\left[
dx\right] \int_0^1\left[ dy\right]\int_0^1\left[ dz\right]
\phi^{\dagger}_{V}\left(z_i\right)
\phi^{\dagger}_p\left(y_i\right)T_{\mathrm
H}\left(x_i,y_i,z_i;\hat{s},\hat{t}\right)
\phi_p\left(x_i\right).\label{eq:convolution} \ee
The elementary scattering process takes place on the quark-gluon
level. Hadrons are replaced by their valence (anti-)quarks which
are assumed to move collinear to their parent hadron with
longitudinal momentum fractions $x_i, y_i, z_i$, respectively. The
internal redistribution of the large transferred momentum is
accomplished by introducing hard gluons connecting all the quark
lines. Thus, the hard scattering amplitude $T_{\mathrm H}$ is a
process dependent, coherent sum of Feynman tree diagrams. For
$T_{\mathrm H}$ we have two generic classes of diagrams: one in
which the photon couples to the vector meson (class I) and one in
which it couples to the proton (class II). Class~I diagrams can be
considered as the remnant of a vector-meson-dominance mechanism,
in which the $\gamma$ fluctuates into the vector meson which then
goes on-shell by exchanging hard gluons with the proton. Class II
diagrams represent a Compton-like mechanism. The arguments
$\hat{s}$ and $\hat{t}$ of $T_{\mathrm H}$ indicate that all quark
masses, apart of the charm-quark mass, are neglected when
calculating $T_{\mathrm H}$. The DAs are probability amplitudes
for the momentum fractions. In our actual calculation we have
chosen the ``asymptotic'' DA for the proton, i.e. $\phi_p \propto
x_1 x_2 x_3$, and the \lq\lq non-relativistic\rq\rq\ DA $\phi_V
\propto \delta\left(z_1- 1/2\right)$ for the vector mesons. For
the strong coupling $\alpha_{\mathrm S}$ we have used the one-loop
expression with $\Lambda_\mathrm{QCD} = 200$ MeV. As argument we
have taken the average virtuality of that hard gluon which
provides a particular $\alpha_{\mathrm S}$. If $|t|$ is large
enough, $\alpha_{\mathrm S}$ should be small enough to justify a
perturbative treatment. In order to extrapolate our predictions to
small $|t|$ values (where data exist), we apply a cutoff such that
$\alpha_{\mathrm S}$ does not exceed a value of 0.7.
\begin{figure}[tb]
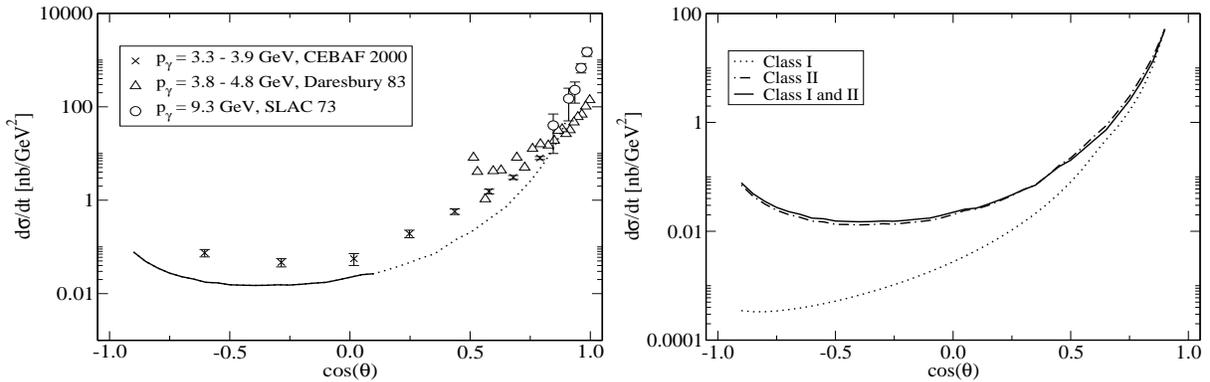

\epsfysize=9.0cm
\begin{center}
\begin{minipage}[t]{8 cm}
\epsfig{file=phi_dsdt_9GeV_lo_Call_asy_new.eps,width=8.0cm, height=5.0cm}
\end{minipage}
\begin{minipage}[t]{8 cm}
\epsfig{file=phi_dsdt_9GeV_lo_CI_CIV_new.eps,width=7.8cm, height=5.0cm}
\end{minipage}
\begin{minipage}[t]{16.5 cm}
\vspace{-0.4cm} \caption{Leading-order perturbative prediction for
$d\sigma_{\gamma p \rightarrow \Phi p}/dt$ at $p_\gamma^\mathrm{
lab}= 9.3$~GeV in comparison with data~\cite{7,8,9} (left figure).
Data from Refs.~\cite{8,9} are appropriately scaled up to
$p_\gamma^\mathrm{lab}= 9.3$~GeV. Results are shown for the
asymptotic proton DA and the non-relativistic DA for the $\Phi$.
The solid part of the curve corresponds to $\alpha_{\mathrm S}\le
0.5$, the remaining part to $0.5 < \alpha_{\mathrm S}\le 0.7$. The
right figure exhibits the contributions of class I and class II
diagrams. \label{fig1}}
\end{minipage}
\end{center}
\end{figure}

Fig.\ref{fig1}~(left) shows our predictions for the unpolarized
$\Phi$-production cross section at $9.3$~GeV photon lab energy in
comparison with experimental data taken at SLAC~\cite{7},
Daresbury~\cite{8} and JLab~\cite{9}. $9.3$~GeV is the highest
energy for which data up to reasonably large scattering angles
exist. To increase our data base we have scaled those data, which
were taken at lower energies but larger scattering angles, such
that they smoothly extrapolate the SLAC data. The angular
dependence of the (scaled) data is well reproduced. The absolute
magnitude is a factor of 3-4 too small, but not orders of
magnitude away as in other photoproduction channels. An increase
of magnitude is even to be expected if one takes the asymptotic DA
for the $\Phi$ ($\phi_V \propto z_1 z_2$) instead of the
non-relativistic one.
Since we have neglected all quark masses for $\Phi$
production, the proton helicity is conserved and the $\Phi$ must
be polarized longitudinally.
The right plot shows how the
differential cross section is composed of class I and class II
contributions. One observes that class II diagrams are by far
dominant. This is somewhat unexpected, since class I graphs
resemble the common picture of vector-meson photoproduction.

Our investigation of $J/\Psi$-photoproduction revealed that it is
crucial to take into account the c-quark mass. The finite c-quark
mass gives rise to the production of transversely polarized
$J/\Psi$s. For reasonably large energies and momentum transfers,
which could be reached, e.g., by an upgrade of CEBAF, the
corresponding amplitudes even dominate nearly over the whole angular range.\\

\noindent {\bf Acknowledgement:} This work was supported via an
agreement for scientific and technological cooperation between
Austria and Croatia (\"OAD project number 19/2004). A.G.
acknowledges the support of the \lq\lq Fonds zur F\"orderung der
wissenschaftlichen Forschung in \"Osterreich\rq\rq\ (project DK
W1203-N08).

\end{document}